\documentclass[aps,prl,twocolumn,groupedaddress,showpacs]{revtex4-1}
\usepackage{graphicx}
\usepackage{epstopdf}
\usepackage{amssymb, amsmath}
\usepackage{multirow}
\usepackage{dcolumn}
\usepackage{bm}

\newcommand\ME[3]      {\langle{{#1}}|{{#2}}|{{#3}}\rangle} 

\newcommand\braket[2]  {\langle{{#1}}|{{#2}}\rangle}
\newcommand\PsiES      {\Psi^*}
\newcommand\psiT      {\psi^*_\mathrm{T}}
\newcommand\ES      {E^*}
\newcommand\Hop        {{\hat{H}}}
\newcommand\Eq[1]      {Eq.~\eqref{#1}}

\begin{document}

\title{Excited state calculations in solids by auxiliary-field quantum Monte Carlo}
\author{Fengjie Ma, Shiwei Zhang, and Henry Krakauer}
\affiliation{Department of Physics, College of William and Mary, Williamsburg, VA 23187}

\begin{abstract}
We present an approach for \emph{ab initio\/} many-body calculations of excited states in solids. Using auxiliary-field quantum Monte Carlo, we introduce an orthogonalization constraint to prevent collapse of the stochastic Slater determinants in the imaginary-time propagation. Trial wave functions from density-functional calculations are used for the constraints. Detailed band structures calculated for standard semiconductors are in good agreement with $GW$ and experimental results. For the challenging ZnO wurtzite structure, we obtain a fundamental band gap of $3.26(16)$\,eV, consistent with experiments.
\end{abstract}

\pacs{
71.15.Qe, 
02.70.Ss, 
71.20.-b, 
71.10.-w  
}

\maketitle

The accurate calculation of excited states in extended systems is a leading challenge in modern electronic structure theory. Density functional theory (DFT), which usually gives good results for ground state properties of materials without strong electron correlation, suffers from the well-known band-gap problem 
even for simple semiconductors \cite{hk,ks,Perdew1983}. The use of hybrid functionals, which incorporate a portion of exact exchange from Hartree-Fock (HF) theory, has led to significant improvements for semiconductors with small to medium gaps, but not for large gap systems \cite{hybrid}. Time-dependent DFT and many-body perturbative approaches have shown considerable promise \cite{Onida2002_RMP}. The $GW$ method is perhaps the simplest and most widely used of the latter and has led to dramatically improved results, largely for simple $sp$ bonded materials  \cite{gw}. It is less successful in materials that are more strongly correlated. In wurtzite structure ZnO, for example, the accuracy of $GW$ quasi-particle excitation energies is still a matter of some controversy, involving many factors such as the choice of pseudopotential, approximate exchange-correlation functional, and choice of plasmon-pole model \cite{zno_gw_louie,zno_ppm,zno_psp}. A general approach which allows accurate calculations of electronic excitations across a wide variety of solids is thus very much in need.

Quantum Monte Carlo (QMC) \cite{dmc,afqmc} is a non-perturbative, many-body computational method which is uniquely capable of scaling up to large system sizes and which in principle does not involve empirical parameters. Ground-state calculations with several flavors of QMC have seen significant recent development \cite{dmc,afqmc,FullCI_QMC}. Although some excited states calculations have also been performed, for example, calculations in molecules \cite{filippi,afqmc_c2} and diffusion Monte Carlo (DMC) calculations of excitation energies of silicon and diamond at some high symmetry $k$-points \cite{dmc_si,dmc_c}, the capability of QMC to treat excited states in general is much less developed. A major reason for this is the intrinsic difficulty of maintaining orthogonality with the lower-lying states when the targeted many-body excited state is being represented stochastically in an imaginary-time projection method. 

The auxiliary-field quantum Monte Carlo (AFQMC) method \cite{afqmc} offers a new framework for addressing this difficulty and doing excited state calculations in solids. The random walks in AFQMC take place in a space of Slater determinants. The single-particle orbitals in each Slater determinant are expressed explicitly in terms of a chosen one-particle basis. Thus, in addition to orthogonalizing the single-particle orbitals with each other (as is done in ground-state calculations), one can orthogonalize them with empty (hole) orbitals to remove contamination in the projection. For example, as an approximate constraint, the occupied orbitals can be orthogonalized with unoccupied orbitals defined by a trial wave function. Recent developments in ground state calculations have demonstrated that the AFQMC framework, with properly formulated sign or phase constraints using trial wave functions, allows sign-problem-free simulations with high accuracy across a wide range of both strongly correlated model Hamiltonians \cite{Chang2008} and real material systems \cite{afqmc,afqmc_atom,fs_nm,afqmc_oxide}. 

In this paper, we show how the AFQMC approach can be formulated to accurately calculate excited states in extended systems. The ground-state method is augmented by an orthogonalization constraint with virtual orbitals to prevent the collapse to lower-lying states in the imaginary-time projection.
Simple trial wave functions from HF or DFT are used for the phaseless and orthogonalization constraints. We illustrate the method by calculating the detailed band structures  in diamond structure silicon and carbon, which are, respectively, small and large band gap semiconductors. We then apply the method to determine the bandgap in ZnO, which has drawn much recent attention and which presents a significant challenge, with high-lying strongly correlated $3d$-bands interacting with $3s$ and $3p$ semicore states.

In AFQMC, we target an excited state of the many-body Hamiltonian $\Hop$ with the imaginary-time projection $|\PsiES(\beta)\rangle\propto e^{-\beta\Hop}{|\psiT\rangle}$, where   $|\psiT\rangle$ is a trial wave function of the excited state, and $\beta\equiv N\Delta \tau$ with $N$ the simulation time step and $\Delta \tau$ the Trotter step size. The wave function at any imaginary-time can be thought of as $|\PsiES(\beta)\rangle=\sum_\phi c_\phi\, |\phi_\uparrow\rangle\otimes |\phi_\downarrow\rangle$. Each Slater determinant in the sum is a random walker, and their distribution gives a statistical representation of the coefficients $c_\phi$. Explicitly, $|\phi_\uparrow\rangle$ has the form \mbox{$(\phi_{1},  \dotsm , \phi_{i},  \dotsm , \phi_{m})$}, where each $\phi_i$ is an orbital (expressed in terms of the chosen single-partle basis) that evolves with $\beta$, and $m$ denotes the number of spin-$\uparrow$ electrons. (Below we will suppress the spin index where possible; the spin-$\downarrow$ component has a similar form.) Thus the AFQMC process resembles the propagation of a population of mean-field solutions in time-dependent external fields. Periodically in the propagation the orbitals within each random walker are orthogonalized with each other to ensure that the signal for fermionic antisymmetry is not lost in numerical noise. This step is also needed in ground-state calculations, and indeed even in mean-field calculations. The excited state energy $\ES$ can be calculated by the mixed estimator \cite{afqmc_c2}:
\begin{equation}
\ES(\beta)= \frac{\ME{\psiT}{\Hop}{\PsiES(\beta)}} 
               {\braket{\psiT}{\PsiES(\beta)}}
        \,,
\label{eq:Emixed}
\end{equation}
which converges to the exact result at $\beta\rightarrow \infty$. If $|\psiT\rangle$ belongs to an irreducible representation of the symmetry group of $\Hop$ different from that of the ground state, \Eq{eq:Emixed} is exact for the lowest excited state of that symmetry. Otherwise, imaginary-time projection is considerably more challenging, since  the propagation will tend to collapse to the ground state (or other lower-lying states) \cite{afqmc_c2}.

The AFQMC formalism, however, provides the ability to prevent the collapse by imposing an additional orthogonality constraint naturally, using virtual orbitals. For a concrete illustration, let us consider a targeted many-body state corresponding to the ``single'' excitation of replacing the $i^{\rm th}$ valence orbital by a conduction orbital labeled by $j$ (thus $j>m$). Each random walker is still an $m$-electron Slater determinant. For the purpose of orthogonalization, however, we will regard it as the extended ordered list \mbox{$(\phi_{1},  \dotsm , \phi_{i-1}, {\bar \phi_{i}}, \phi_{i+1},  \dotsm , \phi_{m}, {\bar \phi_{m+1}},  \dotsm ,{\bar \phi_{j-1}},\phi_{j})$}. Any orbital denoted by `-' is a virtual orbital whose \emph{only} role is in the orthogonalization of the $m$ occupied orbitals. As an approximation, we will use the corresponding orbitals in $|\psi_T\rangle$, i.e., ${\bar \phi_{i}}=\phi^T_{i}$. The choice of $|\psi_T\rangle$ is further discussed below. A Gram-Schmidt orthogonalization on this extended set ensures that the following
orthogonality conditions are obeyed:
i)~$\braket{\bar \phi_i}{\phi_j}=0$; 
ii) $\braket{\bar \phi_i}{\phi_k}=0$ for $k\in[i+1 , m]$;
iii) $\braket{\bar \phi_k}{\phi_j}=0$ for  $k\in[m+1 , j-1]$. 
If the valence state $\bar \phi_i$ is degenerate, its partners are not included in the constraint, which is consistent with ground state AFQMC for an open-shell system. Similarly, any degenerate partners of the conduction state $\phi_j$ are not included. After the Gram-Schmidt step, only the $m$ occupied orbitals  are retained and included in the propagation and measurement, until the next time for orthogonalization when the $\bar \phi$'s are re-inserted and the procedure repeated. The above procedure generalizes straightforwardly if the targeted excited state corresponds to ``double'' excitations or beyond.

We take simple trial wave functions directly from DFT calculations for the phaseless constraint \cite{afqmc} and for the orthogonalization. The starting point for constructing $|\psiT(k)\rangle$ at the selected $k$ point in the Brillouin zone (BZ) is the single Slater determinant DFT wave function for the corresponding ground state, $|\psi_T(k)\rangle$. The orbitals $\phi_{\sigma,i}^T$ are given by the eigenfunctions at $k$ based on a well-converged density integrated over $k$-points. For a given excitation of spin-$\sigma$ from an occupied orbital $i$ to an unoccupied orbital $j$, we replace $\phi_{i,\sigma}^T$ by  $\phi_{j,\sigma}^T$. For insulators, the  ground states are closed shell configurations, so this type of excited state Slater determinant will be spin-contaminated in general. To avoid this, we form a two-determinant singlet wave function: 
\begin{equation}
|\psiT\rangle=(a^\dagger_{j\uparrow}a_{i\uparrow}+a^\dagger_{j\downarrow}a_{i\downarrow})|\psi_T\rangle\, ,
\label{eq:WF_singlet}
\end{equation} 
where $a^\dagger_j$ and $a_i$ are creation and destruction operators for unoccupied (conduction) and occupied (valence) states, respectively, and we assume that the spatial part of the valence and conduction orbitals are spin independent in $|\psi_T\rangle$. In degenerate cases, the trial wave function is constructed by considering all possible promotions among these orbitals, and the coefficients of each determinant are set equal. All our calculations are performed in the primitive cell. In supercells, the folding of bands creates additional mixing of crystal momentum eigenstates,  whose decoupling will need further study. 
 
\begin{figure}
\includegraphics[width=0.42\textwidth]{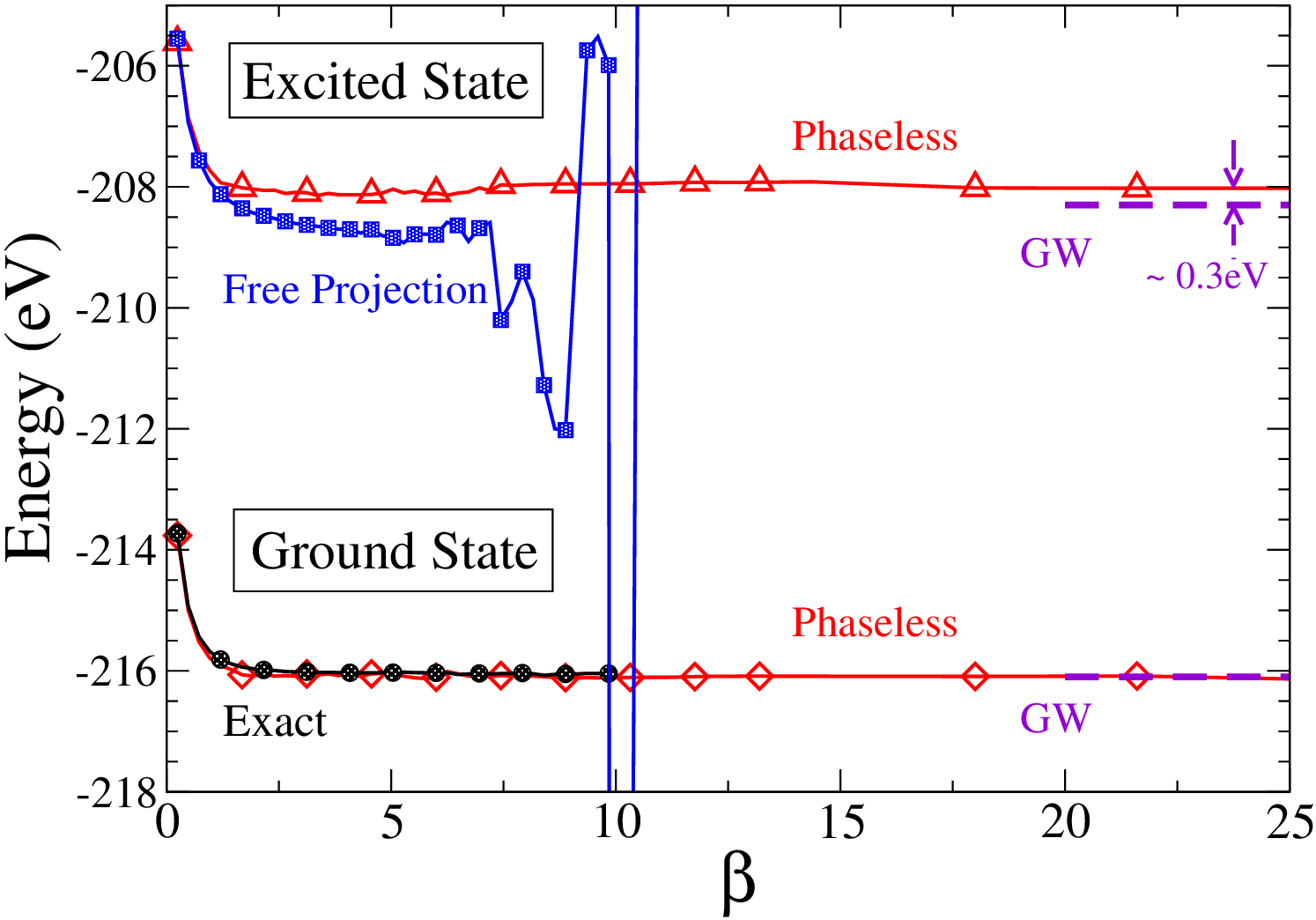}
\caption{(Color online)
Illustration of AFQMC projections of the ground- and excited-state (\mbox{$i=2$ $\to$ $j=5$}) energies [\Eq{eq:Emixed}] in silicon at \mbox{$k=(0.3,0,0)$}. 
Phaseless AFQMC result for the ground state (red diamonds) is compared to exact free-projection (black circles). For the excited state, free-projection (blue squares) collapses. Result from the new method with phaseless and virtual orthogonalization constraint is shown in red triangles, and compared to the
GW excitation, indicated by violet dashed lines \cite{gw-Si}.}
\label{fig:si_freepj}
\end{figure}

Figure~\ref{fig:si_freepj} illustrates our method in fcc silicon. The excited state corresponds to the excitation from band $i=2$ to the lowest lying conduction band $j=5$ at $k=(0.3,0,0)$. The trial wave function $|\psi_T\rangle$ was obtained from DFT with the local-density approximation (LDA), while for excited state  $|\psiT\rangle$ is the singlet trial wave function of \Eq{eq:WF_singlet}. For comparison, AFQMC results are also shown from free-projection, which does not impose the phaseless constraint \cite{afqmc} and is exact for the ground state, but is eventually overwhelmed by the fermion sign problem. (Large runs with $\sim 2\times10^6$ walkers were used in the free-projection calculations.) The ground state phaseless AFQMC results are in excellent agreement with the exact calculation, indicating very small systematic error from the phaseless approximation, consistent with earlier results \cite{betatin_Purwanto}. In contrast with the ground state, exactness is not ensured in free projection for the excited state; a more severe sign problem and onset of collapse to the ground state is seen. The phaseless and the additional orthogonality constraints stabilize the excited state calculation and yield an accurate excitation energy, which is within $\sim 0.3$\,eV of the GW result \cite{gw-Si}  after correction of finite-size effects, as we discuss next. 

Excitation energies from valence state $i$ to conduction state $j$ at any selected BZ $k$ point are calculated as the difference between the AFQMC total energy and the corresponding ground-state energy 
\begin{equation}
\Delta E^\mathrm{QMC}_{ij}(k) = E^*_{ij}(k)-E_0(k).
\label{eq:E_excitation}
\end{equation}
Because the many-body calculations are performed in finite-size (FS) simulation cells, the energies have FS errors which must be removed \cite{Kent1999,Chiesa2006,fs_nm,fs_polar,Drummond2008}. FS corrections to the thermodynamic limit of an infinite-sized supercell can be obtained  as post-processing corrections \cite{fs_nm,fs_polar} from lower-levels of theory for both one-body (1b) and two-body (2b) effects. In the present excited state calculations, however, the creation of the electron-hole (eh) pair results in an additional bias, from the interaction between the particle and hole. We obtain the combined 1b and eh FS correction from DFT calculations:
\begin{equation}
\delta E^\mathrm{eh,1b}_{ij}(k)=e_g(k)- \Delta E^{{\rm DFT}}_{ij}(k)\,, \label{eq:ehole}  
\end{equation}
where \mbox{$e_g(k)\equiv e_j(k)-e_i(k)$} is the difference in band energies from a standard DFT calculation, using a dense $k$-point grid, while $\Delta E^{{\rm DFT}}_{ij}(k)$ is from self-consistent DFT calculations at $k$ paralleling the QMC calculations in Eq.~(\ref{eq:E_excitation}). We correct the 2b FS error, which along with the 1b effect are substantially reduced here because $\Delta E^\mathrm{QMC}_{ij}(k)$ is an energy difference between two states within the same simulation cell, using an LDA functional specially parametrized \cite{fs_nm,fs_polar} for FS calculations
\begin{equation}
\delta E^\mathrm{2b}_{ij}(k)=\Delta E^\mathrm{LDA}_{ij}(k)-\Delta E^\mathrm{FS-LDA}_{ij}(k)\,,
 \label{eq:FS-2b}
\end{equation}
where the two excitation energies on the right are from LDA calculations paralleling the QMC, using the standard and the FS functionals, respectively. The sum of Eqs.~(\ref{eq:ehole}) and (\ref{eq:FS-2b}) gives the total FS correction $\delta E_{ij}(k)$. The largest contribution is from $\delta E^\mathrm{eh,1b}_{ij}(k)$, typically $\sim 0.10$\,eV at most $k$ points in Si and diamond. Its largest value in Si is 0.35\,eV at the $\Gamma$ point. In diamond, which has a large band gap, its largest value is 0.83\,eV. The 2b correction is typically smaller; its largest value is 0.12\,eV in Si and diamond, and  0.08\,eV for the fundamental gap in ZnO.

We then obtain a quasiparticle band structure $\epsilon_n(k)$ from a least-squares fit \citep{dmc_si}  to the calculated many-body excitation energies \mbox{$\Delta E_{ij}(k)\equiv \Delta E^\mathrm{QMC}_{ij}(k) +\delta E_{ij}(k)$}: 
\begin{equation}
\min \left( \sum_{i\in\mathrm{v}}\sum_{j\in\mathrm{c}} \left(\Delta E_{ij}(k)-[\epsilon_j(k)-\epsilon_i(k)]\right)^2 \right)
\label{eq:fit_expression}
\end{equation} 
where $i$ and $j$ run over the occupied (v) and unoccupied (c) states, respectively. The highest occupied quasiparticle energy is
set equal to the corresponding DFT eigenvalue $\epsilon_m(k)=e_m(k)$.

\begin{figure}
\includegraphics[width=0.42\textwidth]{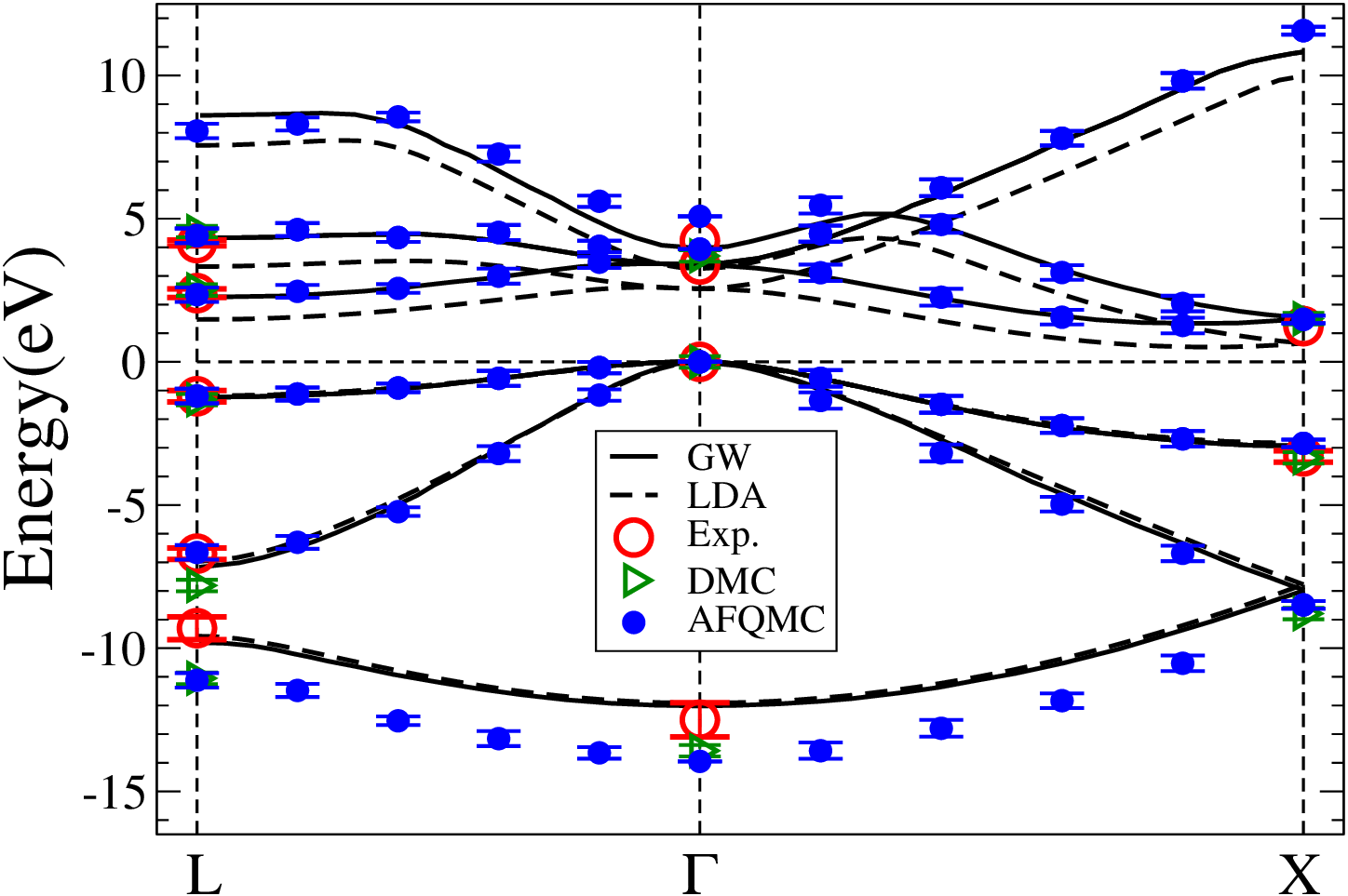}
\caption{(Color online) Many-body band structure of silicon from AFQMC (blue solid circles). GW \cite{gw-Si} and LDA band structures are plotted by solid and dashed black lines, respectively. Available DMC results at high symmetry points, $\Gamma$, $X$, and $L$ \cite{dmc_si}, are indicated by green triangles; experimental values are given by red open circles.}	
\label{fig:si_band}
\end{figure}

The full many-body Si band structure, with FS correction included, is shown in Fig. \ref{fig:si_band} and compared to LDA and $GW$, as well as to available DMC and experimental results \cite{gw-Si,dmc_si}. Calculations for fcc Si were done at the experimental lattice constant of 5.431\,$\AA$. Both AFQMC and DFT calculations used a norm-conserving Kleinman-Bylander type separable nonlocal LDA pseudopotentials generated by the OPIUM code \cite{opium}, with a planewave cutoff $E_\mathrm{cut}=25$\,Ry. Using a $|\psiT\rangle$ from LDA, AFQMC results correct the band gap problem and are in good agreement with experiment and with $GW$. The lowest band, which corresponds to the highest excitation energies, tends to be $\sim 1.5$\,eV too low. For an imaginary-time projection method, its quality can be expected to decline for higher excited states.  Also the simple singlet $|\psiT\rangle$ or the orthogonalization constraint may not be sufficient, as there are many states with similar energies. 

\begin{figure}
\includegraphics[width=0.42\textwidth]{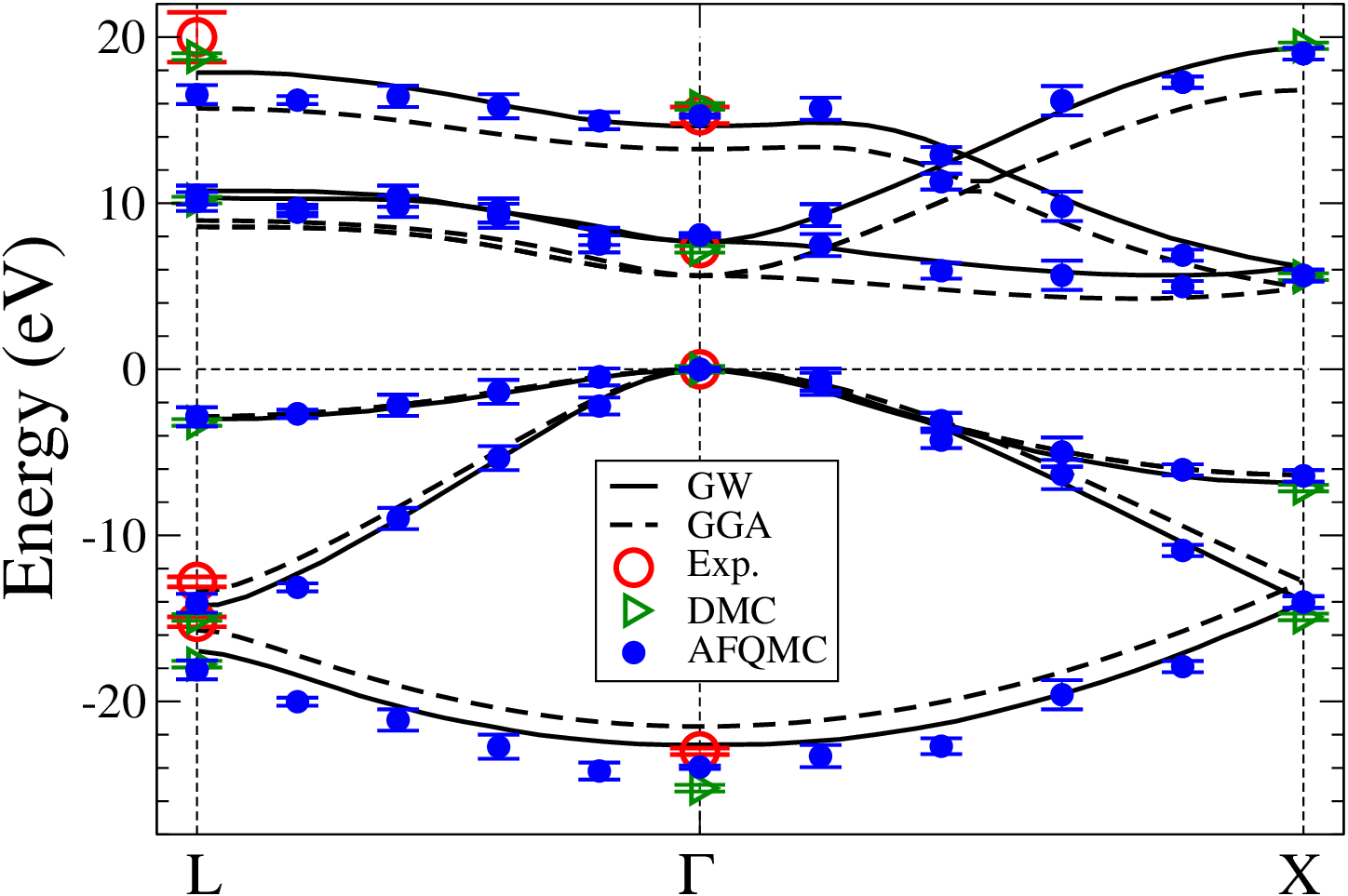}
\caption{(Color online) Many-body band structure of diamond from AFQMC (blue solid circles). GW \cite{gw-C} and GGA band structures are plotted by solid and dashed black lines, respectively. Available DMC results at high symmetry points, $\Gamma$, $X$, and $L$  \cite{dmc_c}, are indicated by green triangles; experimental values are given by red open circles.}
\label{fig:c_band}
\end{figure}

The many body band structure of diamond, with FS correction included, is given in Fig.~\ref{fig:c_band} together with available DMC and experimental results \cite{dmc_c, gw-Si}. The lattice constant of  3.567\,$\AA$ was used. The calculations are similar to those in Si, except for a higher planewave cutoff $E_\mathrm{cut}=50$\,Ry and the use of GGA pseudopotential and trial wave function. We have verified that the calculations are insensitive to the difference in $|\psiT\rangle$ at the level of LDA vs.~GGA or a hybrid functional. The AFQMC results are again generally in very good agreement  with $GW$ and experiment. 

Accurately calculating the fundamental band gap of wurtzite structure ZnO is very challenging. DFT LDA and GGA underestimate the gap by almost $3$\,eV. Even the generally more accurate $GW$ method underestimates the gap by \mbox{$0.8$\,-\,$1.1$\,eV}. There has been considerable discussion about the importance
of various convergence issues, choice of pseudopotentials, and additional approximations such as the plasmon-pole model and the inclusion of self-consistency 
in the $GW$ calculation \cite{zno_gw_louie,zno_ppm,zno_psp}. Results are also sensitive to the choice of pseudopotential, since there is large spatial overlap among $3s$, $3p$ and $3d$ wave functions of atomic Zn \cite{zno_psp}. To properly treat this, our Zn GGA pseudopotential was constructed with a Ne-core, thereby fully correlating the semicore $3s$, $3p$ along with the $3d$ states in the many-body calculation. Very conservative radial cutoffs of $0.83$, $1.02$, and $1.13$~bohr were used for $s$, $p$, and $d$ channels, respectively, resulting in a large planewave cutoff energy of $E_\mathrm{cut}=180$\,Ry. The pseudopotential gives GGA optimized lattice parameters of $a=3.279$ and $c=5.284\,\AA$, and bulk modulus of $128.7$\,GPa, all in good agreement with published GGA results \cite{zno_dft_B}. 

Our AFQMC calculations were done at the above GGA optimized geometry. While hybrid functionals seem to perform better in ZnO, we chose to use simple trial 
wave functions from GGA in our AFQMC calculations to avoid any parameter tuning. The singlet form $|\psiT\rangle$ in Eq.~(\ref{eq:WF_singlet}) was used for the excited state. The main calculations were done with a time-step of $\Delta\tau = 0.01 \mathrm{Ry}^{-1}$. The Trotter error was then corrected by extrapolation from separate runs using a single-determinant $|\psiT\rangle$. The calculated raw band gap was 2.54(14)\,eV. Since DFT severely underestimates the band gap, the FS correction is less straightforward in ZnO. At $k=\Gamma$, for example, the single $k$-point self-consistent GGA calculation would yield a metallic ground state. We extrapolated $\delta E^\mathrm{eh,1b}_{ij}(k)$ for $k$ within $0.1$ of the $\Gamma$ point, along two high symmetry lines. 
In doing so, we assume that the electron-hole effect does not introduce a discontinuity in the band dispersion, which is reasonable since it mainly relates to the simulation cell size. This yielded $\delta E^\mathrm{1b,eh}_{ij}(0)=0.58(08)$\,eV. (Had we used the LDA, which has a smaller gap, $\delta E^\mathrm{1b,eh}_{ij}(0)$ would be about 0.41(04)\,eV.) Ths 2b correction is well-behaved: $\delta E^\mathrm{2b}_{ij}(0)=0.08$\,eV. Adding the FS corrections yields our calculated band gap of $3.20(16)$\,eV.

The experimental equilibrium geometry is at $a=3.250$ and $c=5.207\,\AA$ \cite{Desgreniers1998}. To compare our band gap (for the GGA equilibrium geometry) to experiment, we apply a correction of $+0.06$\,eV, which is the excitation energy difference given by GGA for the experimental and GGA-optimized geometries.
(Hybrid B3LYP calculations give a correction of $+0.10$\,eV.) Table \ref{tab:zno-gap} compares our result with experimental values (3.30 \cite{Srikant1998}, 3.44 \cite{Mang1995}, and 3.57 eV \cite{Tsoi2006,Alawadhi2007}) and those from recent calculations \cite{zno_ppm,zno_blugel,zno_ee,zno_pbe0,zno_hse}. 
We note that, even with the small-core pseudopotential, there can be pseudoization and core-relaxation errors  \cite{Gomez-Abal2008}.  The precise effect cannot be determined without further many-body calculations. However, recent studies \cite{zno_blugel,zno_ee} comparing all-electron and pseudopotential $GW$ calculations indicate that the effect is approximately $+0.27$\,eV for a pseudopotential of similar quality to the one we have adopted. This would increase the
AFQMC result for the  fundamental band gap in ZnO to $\sim 3.53(16)$\,eV, within the range of experimental measurements \cite{Srikant1998,Mang1995,Tsoi2006,Alawadhi2007} listed in Table \ref{tab:zno-gap}.

\begin{table}
\caption{\label{tab:zno-gap} The calculated AFQMC band gap (eV) of wurtzite ZnO,
compared to experiments \cite{Srikant1998,Mang1995,Tsoi2006,Alawadhi2007}. Also shown are results from GGA, hybrid DFT \cite{zno_pbe0,zno_hse}, and $GW$\cite{zno_ppm,zno_blugel,zno_ee}.}
\begin{ruledtabular}
\begin{tabular}{cccccc}
& GGA & Hybrid & GW & AFQMC & Expt. \\
\hline
$\Delta_{gap}$ & 0.77 & 3.32,2.90 & 2.35,2.83,2.56 & 3.26(16) & 3.30-3.57 \\
\end{tabular}
\end{ruledtabular}
\end{table}

Various further improvements can be explored. We have used a planewave basis and norm-conserving pseudopotentials in these calculations. Other single-particle basis sets are straightforward to use, for example, with localized or natural orbitals. One could also work with a truncated set of orbitals from a lower-level of theory, or use a down-folded Hamiltonian directly to improve computational efficiency. The constraining virtual orbitals have been fixed in our calculations, but could potentially be allowed to dynamically evolve in some way in the propagation. It is reassuring that DFT trial wave functions have worked well. Clearly more elaborate multi-determinant wave functions can be used, for example, from diagonalization in a subspace formed by single and double excitations to conduction orbitals.

In summary, we have presented an AFQMC approach for the calculation of electronic excitations in solids, introducing an orbitally-based orthogonalization constraint in the phaseless AFQMC  framework to stabilize the projection of excited states. Simple trial wave functions directly from DFT calculations were used for the constraint. Detailed many-body quasiparticle band structures can be calculated. In prototypical semiconductors (Si and diamond), the calculated band structures are in good agreement with those from $GW$ calculations and with experiment. In the  more strongly correlated and challenging wurtzite ZnO crystal, the calculated  fundamental gap is in excellent agreement with the latest experimental data. The method is non-perturbative and free of empirical parameters, offering a possible path for general computations in correlated materials.

This work is supported by DOE (DE-FG02-09ER16046), NSF (DMR-1006217), and ONR (N000140811235; N000141211042). An award of computer time was provided by the Innovative and Novel Computational Impact on Theory and Experiment (INCITE) program, using resources of the Oak Ridge Leadership Computing Facility at the Oak Ridge National Laboratory, which is supported by the Office of Science of the U.S. Department of Energy under Contract No. DE-AC05-00OR22725.
We also acknowledge the computing support from the Center for Piezoelectrics by Design. We thank W. Purwanto for many useful discussions, and Eric J. Walter for help with pseudopotentials.


\begin{thebibliography}{40}
\expandafter\ifx\csname natexlab\endcsname\relax\def\natexlab#1{#1}\fi
\expandafter\ifx\csname bibnamefont\endcsname\relax
  \def\bibnamefont#1{#1}\fi
\expandafter\ifx\csname bibfnamefont\endcsname\relax
  \def\bibfnamefont#1{#1}\fi
\expandafter\ifx\csname citenamefont\endcsname\relax
  \def\citenamefont#1{#1}\fi
\expandafter\ifx\csname url\endcsname\relax
  \def\url#1{\texttt{#1}}\fi
\expandafter\ifx\csname urlprefix\endcsname\relax\def\urlprefix{URL }\fi
\providecommand{\bibinfo}[2]{#2}
\providecommand{\eprint}[2][]{\url{#2}}

\bibitem[{\citenamefont{Hohenberg and Kohn}(1964)}]{hk}
\bibinfo{author}{\bibfnamefont{P.}~\bibnamefont{Hohenberg}} \bibnamefont{and}
  \bibinfo{author}{\bibfnamefont{W.}~\bibnamefont{Kohn}},
  \bibinfo{journal}{Phys. Rev.} \textbf{\bibinfo{volume}{136}},
  \bibinfo{pages}{B864} (\bibinfo{year}{1964}).

\bibitem[{\citenamefont{Kohn and Sham}(1965)}]{ks}
\bibinfo{author}{\bibfnamefont{W.}~\bibnamefont{Kohn}} \bibnamefont{and}
  \bibinfo{author}{\bibfnamefont{L.~J.} \bibnamefont{Sham}},
  \bibinfo{journal}{Phys. Rev.} \textbf{\bibinfo{volume}{140}},
  \bibinfo{pages}{A1133} (\bibinfo{year}{1965}).

\bibitem[{\citenamefont{Perdew and Levy}(1983)}]{Perdew1983}
\bibinfo{author}{\bibfnamefont{J.~P.} \bibnamefont{Perdew}} \bibnamefont{and}
  \bibinfo{author}{\bibfnamefont{M.}~\bibnamefont{Levy}},
  \bibinfo{journal}{Physical Review Letters} \textbf{\bibinfo{volume}{51}},
  \bibinfo{pages}{1884} (\bibinfo{year}{1983}).

\bibitem[{\citenamefont{Becke}(1993)}]{hybrid}
\bibinfo{author}{\bibfnamefont{A.~D.} \bibnamefont{Becke}},
  \bibinfo{journal}{J. Chem. Phys.} \textbf{\bibinfo{volume}{98}},
  \bibinfo{pages}{1372} (\bibinfo{year}{1993}).

\bibitem[{\citenamefont{Onida and Rubio}(2002)}]{Onida2002_RMP}
\bibinfo{author}{\bibfnamefont{G.}~\bibnamefont{Onida}} \bibnamefont{and}
  \bibinfo{author}{\bibfnamefont{A.}~\bibnamefont{Rubio}},
  \bibinfo{journal}{Rev. Mod. Phys.} \textbf{\bibinfo{volume}{74}},
  \bibinfo{pages}{601} (\bibinfo{year}{2002}).

\bibitem[{\citenamefont{Hedin}(1965)}]{gw}
\bibinfo{author}{\bibfnamefont{L.}~\bibnamefont{Hedin}},
  \bibinfo{journal}{Phys. Rev.} \textbf{\bibinfo{volume}{139}},
  \bibinfo{pages}{A796} (\bibinfo{year}{1965}).

\bibitem[{\citenamefont{Shih et~al.}(2010)\citenamefont{Shih, Xue, Zhang,
  Cohen, and Louie}}]{zno_gw_louie}
\bibinfo{author}{\bibfnamefont{B.-C.} \bibnamefont{Shih}},
  \bibinfo{author}{\bibfnamefont{Y.}~\bibnamefont{Xue}},
  \bibinfo{author}{\bibfnamefont{P.}~\bibnamefont{Zhang}},
  \bibinfo{author}{\bibfnamefont{M.~L.} \bibnamefont{Cohen}}, \bibnamefont{and}
  \bibinfo{author}{\bibfnamefont{S.~G.} \bibnamefont{Louie}},
  \bibinfo{journal}{Phys. Rev. Lett.} \textbf{\bibinfo{volume}{105}},
  \bibinfo{pages}{146401} (\bibinfo{year}{2010}).

\bibitem[{\citenamefont{Stankovski et~al.}(2011)\citenamefont{Stankovski,
  Antonius, Waroquiers, Miglio, Dixit, Sankaran, Giantomassi, Gonze,
  C\^{o}t\'{e}, and Rignanese}}]{zno_ppm}
\bibinfo{author}{\bibfnamefont{M.}~\bibnamefont{Stankovski}},
  \bibinfo{author}{\bibfnamefont{G.}~\bibnamefont{Antonius}},
  \bibinfo{author}{\bibfnamefont{D.}~\bibnamefont{Waroquiers}},
  \bibinfo{author}{\bibfnamefont{A.}~\bibnamefont{Miglio}},
  \bibinfo{author}{\bibfnamefont{H.}~\bibnamefont{Dixit}},
  \bibinfo{author}{\bibfnamefont{K.}~\bibnamefont{Sankaran}},
  \bibinfo{author}{\bibfnamefont{M.}~\bibnamefont{Giantomassi}},
  \bibinfo{author}{\bibfnamefont{X.}~\bibnamefont{Gonze}},
  \bibinfo{author}{\bibfnamefont{M.}~\bibnamefont{C\^{o}t\'{e}}},
  \bibnamefont{and} \bibinfo{author}{\bibfnamefont{G.-M.}
  \bibnamefont{Rignanese}}, \bibinfo{journal}{Phys. Rev. B}
  \textbf{\bibinfo{volume}{84}}, \bibinfo{pages}{241201}
  (\bibinfo{year}{2011}).

\bibitem[{\citenamefont{Partoens et~al.}(2010)\citenamefont{Partoens, Saniz,
  Lamoen, and B}}]{zno_psp}
\bibinfo{author}{\bibfnamefont{H.~D.} \bibnamefont{Partoens}},
  \bibinfo{author}{\bibfnamefont{R.}~\bibnamefont{Saniz}},
  \bibinfo{author}{\bibfnamefont{D.}~\bibnamefont{Lamoen}}, \bibnamefont{and}
  \bibinfo{author}{\bibnamefont{B}}, \bibinfo{journal}{J. Phys.: Condens.
  Matter} \textbf{\bibinfo{volume}{22}}, \bibinfo{pages}{125505}
  (\bibinfo{year}{2010}).

\bibitem[{\citenamefont{Foulkes et~al.}(2001)\citenamefont{Foulkes, Mitas,
  Needs, and Rajagopal}}]{dmc}
\bibinfo{author}{\bibfnamefont{W.~M.~C.} \bibnamefont{Foulkes}},
  \bibinfo{author}{\bibfnamefont{L.}~\bibnamefont{Mitas}},
  \bibinfo{author}{\bibfnamefont{R.~J.} \bibnamefont{Needs}}, \bibnamefont{and}
  \bibinfo{author}{\bibfnamefont{G.}~\bibnamefont{Rajagopal}},
  \bibinfo{journal}{Rev. Mod. Phys.} \textbf{\bibinfo{volume}{73}},
  \bibinfo{pages}{33} (\bibinfo{year}{2001}).

\bibitem[{\citenamefont{Zhang and Krakauer}(2003)}]{afqmc}
\bibinfo{author}{\bibfnamefont{S.}~\bibnamefont{Zhang}} \bibnamefont{and}
  \bibinfo{author}{\bibfnamefont{H.}~\bibnamefont{Krakauer}},
  \bibinfo{journal}{Phys. Rev. Lett.} \textbf{\bibinfo{volume}{90}},
  \bibinfo{pages}{136401} (\bibinfo{year}{2003}).

\bibitem[{\citenamefont{Booth et~al.}(2009)\citenamefont{Booth, Thom, and
  Alavi}}]{FullCI_QMC}
\bibinfo{author}{\bibfnamefont{G.~H.} \bibnamefont{Booth}},
  \bibinfo{author}{\bibfnamefont{A.~J.~W.} \bibnamefont{Thom}},
  \bibnamefont{and} \bibinfo{author}{\bibfnamefont{A.}~\bibnamefont{Alavi}},
  \bibinfo{journal}{The Journal of Chemical Physics}
  \textbf{\bibinfo{volume}{131}}, \bibinfo{pages}{54106}
  (\bibinfo{year}{2009}).

\bibitem[{\citenamefont{Schautz et~al.}(2004)\citenamefont{Schautz, Buda, and
  Filippi}}]{filippi}
\bibinfo{author}{\bibfnamefont{F.}~\bibnamefont{Schautz}},
  \bibinfo{author}{\bibfnamefont{F.}~\bibnamefont{Buda}}, \bibnamefont{and}
  \bibinfo{author}{\bibfnamefont{C.}~\bibnamefont{Filippi}},
  \bibinfo{journal}{The Journal of Chemical Physics}
  \textbf{\bibinfo{volume}{121}}, \bibinfo{pages}{5836} (\bibinfo{year}{2004}).

\bibitem[{\citenamefont{Purwanto
  et~al.}(2009{\natexlab{a}})\citenamefont{Purwanto, Zhang, and
  Krakauer}}]{afqmc_c2}
\bibinfo{author}{\bibfnamefont{W.}~\bibnamefont{Purwanto}},
  \bibinfo{author}{\bibfnamefont{S.}~\bibnamefont{Zhang}}, \bibnamefont{and}
  \bibinfo{author}{\bibfnamefont{H.}~\bibnamefont{Krakauer}},
  \bibinfo{journal}{J. Chem. Phys.} \textbf{\bibinfo{volume}{130}},
  \bibinfo{pages}{94107} (\bibinfo{year}{2009}{\natexlab{a}}).

\bibitem[{\citenamefont{Williamson et~al.}(1998)\citenamefont{Williamson, Hood,
  Needs, and Rajagopal}}]{dmc_si}
\bibinfo{author}{\bibfnamefont{A.~J.} \bibnamefont{Williamson}},
  \bibinfo{author}{\bibfnamefont{R.~Q.} \bibnamefont{Hood}},
  \bibinfo{author}{\bibfnamefont{R.~J.} \bibnamefont{Needs}}, \bibnamefont{and}
  \bibinfo{author}{\bibfnamefont{G.}~\bibnamefont{Rajagopal}},
  \bibinfo{journal}{Phys. Rev. B} \textbf{\bibinfo{volume}{57}},
  \bibinfo{pages}{12140} (\bibinfo{year}{1998}).

\bibitem[{\citenamefont{Towler et~al.}(2000)\citenamefont{Towler, Hood, and
  Needs}}]{dmc_c}
\bibinfo{author}{\bibfnamefont{M.~D.} \bibnamefont{Towler}},
  \bibinfo{author}{\bibfnamefont{R.~Q.} \bibnamefont{Hood}}, \bibnamefont{and}
  \bibinfo{author}{\bibfnamefont{R.~J.} \bibnamefont{Needs}},
  \bibinfo{journal}{Phys. Rev. B} \textbf{\bibinfo{volume}{62}},
  \bibinfo{pages}{2330} (\bibinfo{year}{2000}).

\bibitem[{\citenamefont{Chang and Zhang}(2008)}]{Chang2008}
\bibinfo{author}{\bibfnamefont{C.-C.} \bibnamefont{Chang}} \bibnamefont{and}
  \bibinfo{author}{\bibfnamefont{S.}~\bibnamefont{Zhang}},
  \bibinfo{journal}{Physical Review B} \textbf{\bibinfo{volume}{78}},
  \bibinfo{pages}{165101} (\bibinfo{year}{2008}).

\bibitem[{\citenamefont{Suewattana et~al.}(2007)\citenamefont{Suewattana,
  Purwanto, Zhang, Krakauer, and Walter}}]{afqmc_atom}
\bibinfo{author}{\bibfnamefont{M.}~\bibnamefont{Suewattana}},
  \bibinfo{author}{\bibfnamefont{W.}~\bibnamefont{Purwanto}},
  \bibinfo{author}{\bibfnamefont{S.}~\bibnamefont{Zhang}},
  \bibinfo{author}{\bibfnamefont{H.}~\bibnamefont{Krakauer}}, \bibnamefont{and}
  \bibinfo{author}{\bibfnamefont{E.~J.} \bibnamefont{Walter}},
  \bibinfo{journal}{Phys. Rev. B} \textbf{\bibinfo{volume}{75}},
  \bibinfo{pages}{245123} (\bibinfo{year}{2007}).

\bibitem[{\citenamefont{Kwee et~al.}(2008)\citenamefont{Kwee, Zhang, and
  Krakauer}}]{fs_nm}
\bibinfo{author}{\bibfnamefont{H.}~\bibnamefont{Kwee}},
  \bibinfo{author}{\bibfnamefont{S.}~\bibnamefont{Zhang}}, \bibnamefont{and}
  \bibinfo{author}{\bibfnamefont{H.}~\bibnamefont{Krakauer}},
  \bibinfo{journal}{Phys. Rev. Lett.} \textbf{\bibinfo{volume}{100}},
  \bibinfo{pages}{126404} (\bibinfo{year}{2008}).

\bibitem[{\citenamefont{Al-Saidi et~al.}(2006)\citenamefont{Al-Saidi, Krakauer,
  and Zhang}}]{afqmc_oxide}
\bibinfo{author}{\bibfnamefont{W.~A.} \bibnamefont{Al-Saidi}},
  \bibinfo{author}{\bibfnamefont{H.}~\bibnamefont{Krakauer}}, \bibnamefont{and}
  \bibinfo{author}{\bibfnamefont{S.}~\bibnamefont{Zhang}},
  \bibinfo{journal}{Phys. Rev. B} \textbf{\bibinfo{volume}{73}},
  \bibinfo{pages}{075103} (\bibinfo{year}{2006}).

\bibitem[{\citenamefont{Rohlfing et~al.}(1993)\citenamefont{Rohlfing,
  Kr\"{u}ger, and Pollmann}}]{gw-Si}
\bibinfo{author}{\bibfnamefont{M.}~\bibnamefont{Rohlfing}},
  \bibinfo{author}{\bibfnamefont{P.}~\bibnamefont{Kr\"{u}ger}},
  \bibnamefont{and} \bibinfo{author}{\bibfnamefont{J.}~\bibnamefont{Pollmann}},
  \bibinfo{journal}{Phys. Rev. B} \textbf{\bibinfo{volume}{48}},
  \bibinfo{pages}{17791} (\bibinfo{year}{1993}).

\bibitem[{\citenamefont{Purwanto
  et~al.}(2009{\natexlab{b}})\citenamefont{Purwanto, Krakauer, and
  Zhang}}]{betatin_Purwanto}
\bibinfo{author}{\bibfnamefont{W.}~\bibnamefont{Purwanto}},
  \bibinfo{author}{\bibfnamefont{H.}~\bibnamefont{Krakauer}}, \bibnamefont{and}
  \bibinfo{author}{\bibfnamefont{S.}~\bibnamefont{Zhang}},
  \bibinfo{journal}{Physical Review B} \textbf{\bibinfo{volume}{80}},
  \bibinfo{pages}{214116} (\bibinfo{year}{2009}{\natexlab{b}}).

\bibitem[{\citenamefont{Kent et~al.}(1999)\citenamefont{Kent, Hood, Williamson,
  Needs, Foulkes, and Rajagopal}}]{Kent1999}
\bibinfo{author}{\bibfnamefont{P.~R.~C.} \bibnamefont{Kent}},
  \bibinfo{author}{\bibfnamefont{R.~Q.} \bibnamefont{Hood}},
  \bibinfo{author}{\bibfnamefont{A.~J.} \bibnamefont{Williamson}},
  \bibinfo{author}{\bibfnamefont{R.~J.} \bibnamefont{Needs}},
  \bibinfo{author}{\bibfnamefont{W.~M.~C.} \bibnamefont{Foulkes}},
  \bibnamefont{and}
  \bibinfo{author}{\bibfnamefont{G.}~\bibnamefont{Rajagopal}},
  \bibinfo{journal}{Physical Review B} \textbf{\bibinfo{volume}{59}},
  \bibinfo{pages}{1917} (\bibinfo{year}{1999}).

\bibitem[{\citenamefont{Chiesa et~al.}(2006)\citenamefont{Chiesa, Ceperley,
  Martin, and Holzmann}}]{Chiesa2006}
\bibinfo{author}{\bibfnamefont{S.}~\bibnamefont{Chiesa}},
  \bibinfo{author}{\bibfnamefont{D. M.}~\bibnamefont{Ceperley}},
  \bibinfo{author}{\bibfnamefont{R. M.}~\bibnamefont{Martin}}, \bibnamefont{and}
  \bibinfo{author}{\bibfnamefont{M.}~\bibnamefont{Holzmann}},
  \bibinfo{journal}{Physical Review Letters} \textbf{\bibinfo{volume}{97}},
  \bibinfo{pages}{076404} (\bibinfo{year}{2006}).

\bibitem[{\citenamefont{Ma et~al.}(2011)\citenamefont{Ma, Zhang, and
  Krakauer}}]{fs_polar}
\bibinfo{author}{\bibfnamefont{F.}~\bibnamefont{Ma}},
  \bibinfo{author}{\bibfnamefont{S.}~\bibnamefont{Zhang}}, \bibnamefont{and}
  \bibinfo{author}{\bibfnamefont{H.}~\bibnamefont{Krakauer}},
  \bibinfo{journal}{Phys. Rev. B} \textbf{\bibinfo{volume}{84}},
  \bibinfo{pages}{155130} (\bibinfo{year}{2011}).

\bibitem[{\citenamefont{Drummond et~al.}(2008)\citenamefont{Drummond, Needs,
  Sorouri, and Foulkes}}]{Drummond2008}
\bibinfo{author}{\bibfnamefont{N.~D.} \bibnamefont{Drummond}},
  \bibinfo{author}{\bibfnamefont{R.~J.} \bibnamefont{Needs}},
  \bibinfo{author}{\bibfnamefont{A.}~\bibnamefont{Sorouri}}, \bibnamefont{and}
  \bibinfo{author}{\bibfnamefont{W.~M.~C.} \bibnamefont{Foulkes}},
  \bibinfo{journal}{Physical Review B} \textbf{\bibinfo{volume}{78}},
  \bibinfo{pages}{125106} (\bibinfo{year}{2008}).

\bibitem[{opi()}]{opium}
\emph{\bibinfo{title}{{The OPIUM project, available at
  http://opium.sourceforge.net}}}.

\bibitem[{\citenamefont{Faleev et~al.}(2006)\citenamefont{Faleev, van
  Schilfgaarde, Kotani, L\'{e}onard, and Desjarlais}}]{gw-C}
\bibinfo{author}{\bibfnamefont{S.~V.} \bibnamefont{Faleev}},
  \bibinfo{author}{\bibfnamefont{M.}~\bibnamefont{van Schilfgaarde}},
  \bibinfo{author}{\bibfnamefont{T.}~\bibnamefont{Kotani}},
  \bibinfo{author}{\bibfnamefont{F.}~\bibnamefont{L\'{e}onard}},
  \bibnamefont{and} \bibinfo{author}{\bibfnamefont{M.~P.}
  \bibnamefont{Desjarlais}}, \bibinfo{journal}{Phys. Rev. B}
  \textbf{\bibinfo{volume}{74}}, \bibinfo{pages}{033101} (\bibinfo{year}{2006}).

\bibitem[{\citenamefont{Molepo and Joubert}(2011)}]{zno_dft_B}
\bibinfo{author}{\bibfnamefont{M.~P.} \bibnamefont{Molepo}} \bibnamefont{and}
  \bibinfo{author}{\bibfnamefont{D.~P.} \bibnamefont{Joubert}},
  \bibinfo{journal}{Phys. Rev. B} \textbf{\bibinfo{volume}{84}},
  \bibinfo{pages}{094110} (\bibinfo{year}{2011}).

\bibitem[{\citenamefont{Desgreniers}(1998)}]{Desgreniers1998}
\bibinfo{author}{\bibfnamefont{S.}~\bibnamefont{Desgreniers}},
  \bibinfo{journal}{Physical Review B} \textbf{\bibinfo{volume}{58}},
  \bibinfo{pages}{14102} (\bibinfo{year}{1998}).

\bibitem[{\citenamefont{Srikant and Clarke}(1998)}]{Srikant1998}
\bibinfo{author}{\bibfnamefont{V.}~\bibnamefont{Srikant}} \bibnamefont{and}
  \bibinfo{author}{\bibfnamefont{D.~R.} \bibnamefont{Clarke}},
  \bibinfo{journal}{Journal of Applied Physics} \textbf{\bibinfo{volume}{83}},
  \bibinfo{pages}{5447} (\bibinfo{year}{1998}).

\bibitem[{\citenamefont{Mang et~al.}(1995)\citenamefont{Mang, Reimann, and
  R\"{u}benacke}}]{Mang1995}
\bibinfo{author}{\bibfnamefont{A.}~\bibnamefont{Mang}},
  \bibinfo{author}{\bibfnamefont{K.}~\bibnamefont{Reimann}}, \bibnamefont{and}
  \bibinfo{author}{\bibfnamefont{S.}~\bibnamefont{R\"{u}benacke}},
  \bibinfo{journal}{Solid State Communications} \textbf{\bibinfo{volume}{94}},
  \bibinfo{pages}{251} (\bibinfo{year}{1995}).

\bibitem[{\citenamefont{Tsoi et~al.}(2006)\citenamefont{Tsoi, Lu, Ramdas,
  Alawadhi, Grimsditch, Cardona, and Lauck}}]{Tsoi2006}
\bibinfo{author}{\bibfnamefont{S.}~\bibnamefont{Tsoi}},
  \bibinfo{author}{\bibfnamefont{X.}~\bibnamefont{Lu}},
  \bibinfo{author}{\bibfnamefont{A.~K.} \bibnamefont{Ramdas}},
  \bibinfo{author}{\bibfnamefont{H.}~\bibnamefont{Alawadhi}},
  \bibinfo{author}{\bibfnamefont{M.}~\bibnamefont{Grimsditch}},
  \bibinfo{author}{\bibfnamefont{M.}~\bibnamefont{Cardona}}, \bibnamefont{and}
  \bibinfo{author}{\bibfnamefont{R.}~\bibnamefont{Lauck}},
  \bibinfo{journal}{Physical Review B} \textbf{\bibinfo{volume}{74}},
  \bibinfo{pages}{165203} (\bibinfo{year}{2006}).

\bibitem[{\citenamefont{Alawadhi et~al.}(2007)\citenamefont{Alawadhi, Tsoi, Lu,
  Ramdas, Grimsditch, Cardona, and Lauck}}]{Alawadhi2007}
\bibinfo{author}{\bibfnamefont{H.}~\bibnamefont{Alawadhi}},
  \bibinfo{author}{\bibfnamefont{S.}~\bibnamefont{Tsoi}},
  \bibinfo{author}{\bibfnamefont{X.}~\bibnamefont{Lu}},
  \bibinfo{author}{\bibfnamefont{A.~K.} \bibnamefont{Ramdas}},
  \bibinfo{author}{\bibfnamefont{M.}~\bibnamefont{Grimsditch}},
  \bibinfo{author}{\bibfnamefont{M.}~\bibnamefont{Cardona}}, \bibnamefont{and}
  \bibinfo{author}{\bibfnamefont{R.}~\bibnamefont{Lauck}},
  \bibinfo{journal}{Physical Review B} \textbf{\bibinfo{volume}{75}},
  \bibinfo{pages}{205207} (\bibinfo{year}{2007}).

\bibitem[{\citenamefont{Friedrich
  et~al.}(2011{\natexlab{a}})\citenamefont{Friedrich, M\"{u}ller, and
  Bl\"{u}gel}}]{zno_blugel}
\bibinfo{author}{\bibfnamefont{C.}~\bibnamefont{Friedrich}},
  \bibinfo{author}{\bibfnamefont{M.~C.} \bibnamefont{M\"{u}ller}},
  \bibnamefont{and}
  \bibinfo{author}{\bibfnamefont{S.}~\bibnamefont{Bl\"{u}gel}},
  \bibinfo{journal}{Physical Review B} \textbf{\bibinfo{volume}{83}},
  \bibinfo{pages}{081101} (\bibinfo{year}{2011}{\natexlab{b}});
  \textbf{\bibinfo{volume}{84}},
  \bibinfo{pages}{039906(E)} (\bibinfo{year}{2011}{\natexlab{a}}).

\bibitem[{\citenamefont{Berger et~al.}(2012)\citenamefont{Berger, Reining, and
  Sottile}}]{zno_ee}
\bibinfo{author}{\bibfnamefont{J.~A.} \bibnamefont{Berger}},
  \bibinfo{author}{\bibfnamefont{L.}~\bibnamefont{Reining}}, \bibnamefont{and}
  \bibinfo{author}{\bibfnamefont{F.}~\bibnamefont{Sottile}},
  \bibinfo{journal}{Physical Review B} \textbf{\bibinfo{volume}{85}},
  \bibinfo{pages}{085126} (\bibinfo{year}{2012}).

\bibitem[{\citenamefont{Betzinger et~al.}(2010)\citenamefont{Betzinger,
  Friedrich, and Bl\"{u}gel}}]{zno_pbe0}
\bibinfo{author}{\bibfnamefont{M.}~\bibnamefont{Betzinger}},
  \bibinfo{author}{\bibfnamefont{C.}~\bibnamefont{Friedrich}},
  \bibnamefont{and}
  \bibinfo{author}{\bibfnamefont{S.}~\bibnamefont{Bl\"{u}gel}},
  \bibinfo{journal}{Phys. Rev. B} \textbf{\bibinfo{volume}{81}},
  \bibinfo{pages}{195117} (\bibinfo{year}{2010}).

\bibitem[{\citenamefont{Uddin and Scuseria}(2006)}]{zno_hse}
\bibinfo{author}{\bibfnamefont{J.}~\bibnamefont{Uddin}} \bibnamefont{and}
  \bibinfo{author}{\bibfnamefont{G.~E.} \bibnamefont{Scuseria}},
  \bibinfo{journal}{Phys. Rev. B} \textbf{\bibinfo{volume}{74}},
  \bibinfo{pages}{245115} (\bibinfo{year}{2006}).

\bibitem[{\citenamefont{G\'{o}mez-Abal
  et~al.}(2008)\citenamefont{G\'{o}mez-Abal, Li, Scheffler, and
  Ambrosch-Draxl}}]{Gomez-Abal2008}
\bibinfo{author}{\bibfnamefont{R.}~\bibnamefont{G\'{o}mez-Abal}},
  \bibinfo{author}{\bibfnamefont{X.}~\bibnamefont{Li}},
  \bibinfo{author}{\bibfnamefont{M.}~\bibnamefont{Scheffler}},
  \bibnamefont{and}
  \bibinfo{author}{\bibfnamefont{C.}~\bibnamefont{Ambrosch-Draxl}},
  \bibinfo{journal}{Physical Review Letters} \textbf{\bibinfo{volume}{101}},
  \bibinfo{pages}{106404} (\bibinfo{year}{2008}).

\end{thebibliography}
\end{document}